\documentclass[sigconf]{acmart}
\usepackage[utf8]{inputenc}
\usepackage[T1]{fontenc}
\usepackage[english]{babel}

\title{On the need for effective tools for debugging quantum programs}

\author{Olivia Di Matteo}
\email{olivia@ece.ubc.ca}
\orcid{0000-0002-1372-7706}
\affiliation{%
  \institution{Department of Electrical and Computer Engineering, The University of British Columbia}
  \streetaddress{2332 Main Mall}
  \city{Vancouver}
  \state{British Columbia}
  \country{Canada}
  \postcode{V6T 1Z4}
}

\copyrightyear{2024} 
\acmYear{2024} 
\setcopyright{acmlicensed}\acmConference[Q-SE 2024]{2024 ACM/IEEE International Workshop on
Quantum Software Engineering }{April 16, 2024}{Lisbon, Portugal}
\acmBooktitle{2024 ACM/IEEE International Workshop on Quantum Software Engineering (Q-SE 2024), April
16, 2024, Lisbon, Portugal}
\acmDOI{10.1145/xxxxxxx.xxxxxxx}

\acmSubmissionID{ICSE-WORKSHOP-QSE-13}

\begin{document}

\begin{abstract}
The ability to incorporate quantum phenomena in computing unlocks a host of new ways to make mistakes. This work surveys existing studies and approaches to debugging quantum programs. It then presents a set of examples that stem from first-hand experience, intended to motivate future research on the subject and the development of novel tools and techniques. 
\end{abstract}

\maketitle

\section{Introduction}
Debugging is an essential skill for both developers and users of software. Quantum computing (QC), where unintuitive phenomena affect the conceptualization and development of algorithms, necessitates specialized debugging techniques. However, there is a relative dearth of guidance and tools for detecting, isolating, and fixing quantum bugs. 

We begin by making a distinction between bugs in  (a) the implementation of a quantum programming language or framework, (b) quantum algorithms implemented using these frameworks; our focus is the latter. More explicitly, we define \emph{quantum bugs} as bugs that occur precisely
\emph{because} an algorithm is quantum, and thus require domain-specific
knowledge to fix. Incorrect gate sequences, mishandling of measurement processes, and the need to integrate disparate tools and conventions can lead to unexpected behaviour which is challenging to detect and diagnose, especially when the probabilistic nature of QC means they may not always manifest in the same way. Detecting bugs in a hardware setting is further complicated by noise that may exacerbate, or worse, conceal them.

Debugging tactics from classical software engineering such as backtracking, runtime data analysis, and testing, serve as a first line of defense \cite{miranskyy2020bugfree, miranskyy2021testing, zhao2021horizons, garcia2023quantum, long2023testing}. Most quantum frameworks already perform unit and integration testing, but even well-structured tests, written by developers with deep knowledge of QC, can miss edge cases. Furthermore, experienced developers cannot anticipate the infinite number of ways in which users new to QC (currently, most users) will do something unexpected and uncover a bug.

The lack of integration between contemporary quantum programming tools has also led to a software ecosystem in which bugs thrive. Developers often spend a significant amount of time as ``quantum plumbers'', building pipes to connect frameworks with different qubit orderings, gate sets, level of documentation detail, test conventions, and programming styles. Even if an algorithm is initially correct, this conversion may introduce new problems.

In this work we discuss the need for more effective tools for debugging quantum programs.  
We begin with a brief overview of existing work: what kinds of quantum bugs  occur, how are they diagnosed, and what structures are in place to prevent them?  We then present examples of real bugs in quantum programs. These motivate a set of research questions, put forward as a call for action and collaboration within the quantum software community.

\section{Existing work}

\subsection{Manifestations of quantum bugs}

To develop concrete debugging strategies it is useful to study a variety of real examples \cite{campos2021qbugs}. A number of empirical studies analyzed bugs in open-source repositories.
Ref. \cite{zhao2021bugs4q} catalogued over 200 bugs from Qiskit, finding most resulted in incorrect program output. Ref. \cite{zhao2021identifying} identified \emph{bug patterns} developing algorithms in Qiskit, and suggested fixes and preventative measures. Ref. \cite{paltenghi2022bugs} investigated 18 frameworks and found $\approx$40\% of bugs were ``quantum-specific'', and identified where such bugs were most likely to occur (notably, circuit optimization routines). 
\cite{zhao2023empiricawl} studied bugs in 22 quantum machine learning projects; 28\% were quantum-specific, and exceptions, followed by function errors, were the most frequent symptoms.
In \cite{luo2022comprehensive} bugs encountered using Qiskit, Cirq, Q\#, and ProjectQ were analyzed and over $80\%$ deemed quantum-specific. Most required single-line fixes, and they concluded ``this implies that the cost of fixing bugs is generally low in existing quantum programs''. However, many bugs pertained to, e.g., API usage, and are not quantum bugs as defined herein, even if the fix required domain knowledge.  

In addition to the previous point, these studies have other limitations.
As the focus is primarily on frameworks, they are not always representative of how quantum programmers \emph{use} them. 
Furthermore, many focus on Qiskit, limiting the scope to one user base and programming style (object-oriented gate-model algorithms).
However, analyzing these bugs uncovers common features that improve our understanding.

\subsection{Quantum bug diagnosis: methods and tools}

A common approach to debugging is breakpoints and assertions. However, when executing a quantum algorithm on hardware, one cannot pause to inspect the state. On a simulator, parsing large state vectors can be unmanageable, motivating tools with easy-to-read outputs. To that end, assertion-based debugging was explored in \cite{huang2019statistical}, motivated by a bug taxonomy and case studies. \cite{huang2019statistical} proposes statistical assertions to test if a quantum register is in a classical state, uniform superposition, or entangled state. Property-testing strategies were developed in \cite{honarvar2020propertybased} for Q\# programs. However, validating such assertions requires halting a program for measurement. Ideally, states could be copied and tested offline, but the no-cloning theorem prohibits this. Workarounds include \emph{approximate cloning} \cite{wang2023debugcloning} or measuring auxiliary qubits \cite{li2020proq, liu2020dynamic}.

Interactive debugging tools have also been developed. Most programming frameworks offer circuit visualization and ways to ``snapshot'' programs during execution, which are useful for validating small algorithms.
Quirk \cite{quirk} is a web-based circuit composer that includes real-time diagnostic information.
QChecker performs static analysis to detect bugs in Qiskit programs \cite{qchecker}. Cirquo \cite{metwalli2023cirquo} enables users to debug Qiskit circuits by adding instructions to slice circuits into subcircuits to analyze individually. Quantivine \cite{wen23_quant}, a  visualization tool for large-scale circuits, can ``zoom in" on subroutines to identify where an error occurs. Standard debugging tools in Visual Studio can be used with Q\#, which itself has functionality for dumping machine state, assertions, and developing unit tests \cite{qsharp}. The Qiskit Trebugger \cite{trebugger} (transpiler debugger) logs and displays transpiler passes applied to circuits during optimization. As with the empirical studies, many tools are tied to a specific framework, though their strategies can in principle be applied more broadly. 

\subsection{Prevention strategies}
At the framework level testing is essential  (see \cite{garcia2023quantum} for a recent review of quantum software testing). Here we focus on what can be done by and for users implementing algorithms.
Methods such as correctness-by-construction \cite{peduri2023qbc} and formal verification (FV) \cite{ying2019toward, chareton2022formal, lewis2023formal} can validate the behaviour of quantum programs.
\cite{chareton2022formal} distinguishes between \emph{low-level verification} like equivalence checking (EC), and \emph{high-level verification} that certifies an entire algorithm is correct. 
 EC is critical in compilation, which should modify programs without changing their behaviour. The simplest means of EC is comparing matrices, but this has obvious scaling issues. Other approaches include path sums \cite{amy2019towards}, decision diagrams \cite{burgholzer2021qcec, hong2022equivalence, peham2022equivalenceparadigms}, ZX-calculus \cite{peham2022equivalence, peham2023equivalence}, and simulative techniques \cite{burgholzer2020random}.

EC answers the question "are these circuits the same?", while FV addresses "does this program provably work as intended?".
FV methods are based on mathematical models and usually rely on an automated theorem prover or proof assistant, directed by a user. Quantum aspects are typically embedded in classical frameworks to leverage functionality such as symbolic computation, an approach taken by SQIR \cite{hietala2021proving} and QWIRE \cite{qwire}. Quantum Hoare Logic performs FV using density matrices \cite{ying2019toward}, and
\cite{qbricks} and \cite{amy2019formalphd} manipulate operations as path-sums. FV can also be applied to compilation: VOQC \cite{hietala2021voqc} verifies circuit optimization and qubit mapping, and CertiQ \cite{shi2020certiq} automatically verifies most Qiskit transpiler passes.

The reader may be wondering ``if FV certifies correctness, why do we even need debugging tools?''. FV has primarily been applied to textbook algorithms with closed-form solutions, and it is less clear how it applies to new ones.
There is also a lack of tool integration; mathematical proof assistants are a barrier for quantum programmers used to Python frameworks, but automation will make it more accessible \cite{ying2019toward}.
Finally, while FV can prove a program is correct, \emph{it does not tell you why it is wrong}. A classical compiler will flag missing semicolons; current FV tools can't flag missing Hadamards. 

A final prevention strategy is design of quantum programming languages. Functional approaches with deliberate types system are found in early languages such as QPL \cite{selinger2004towards} and cQPL \cite{Mauerer2005SemanticsAS}, and recent ones \cite{hietala2021proving, qwire, qsharp, protoquipperm, silq}, to ensure tenets such as no-cloning are upheld. Some languages are also equipped with features to decrease the load of the user. For instance, Silq's \cite{silq} type system determines which parts of a program can be safely (and automatically) uncomputed. The abstraction level of frameworks is also improving, and automating construction of common subroutines should reduce the amount of errors involving, e.g., incorrect qubit indexing. 

\section{Examples}

To develop effective debugging strategies, there is no substitute for first-hand experience.
The following bugs were identified (i.e., caused) by the author while 
implementing a variational quantum eigensolver (VQE) \cite{vqe} and running it on hardware \cite{oxygen}.
One can argue that the author should just ``be better at quantum programming'', or ``not make so many mistakes''. But, everyone makes mistakes, and we all benefit from better debugging tools.
The focus here is on \emph{quantum bugs} that occur due to quantum phenomena, and whose debugging process is complicated by them. The first bug is conceptual, while the others relate more to usage of frameworks.

\subsection{Example 1: when global phase \emph{does} matter}

\emph{Context:} The initial VQE circuits were expressed using quantum gates at a high level of abstraction. To run on hardware, compilation into 1- and 2-qubit gates that leveraged shortcuts and custom decompositions was performed. \emph{Symptom:} The output of the VQE after manual circuit optimization was off by a few decimal points.

\emph{Analysis}. While programming frameworks include default decompositions for high-level gates, it is sometimes useful to override them. The VQE circuits included Toffolis and controlled Toffolis, for which decompositions with different circuit depths and gate counts exist. Toffoli decompositions are typically given in the $\{H, T, CNOT\}$ gate set, while we were using $\{RY, RZ, CNOT\}$. As a shortcut, $RZ(\pi/4)$ was used instead of $T$, yielding a Toffoli up to a global phase.
However, with the new decomposition the algorithm no longer produced correct results. VQE optimization was initially thought to be the issue, so parameters were reoptimized and convergence tested. EC was then performed, surfacing a discrepancy in matrix representations: the global phase led to relative phases when the Toffoli was applied controlled on other qubits. 
An obvious prevention strategy is to not prematurely optimize, but frameworks could also help by, e.g., including EC for custom decompositions that raises warnings if equivalence is up to a global phase.   

\subsection{Example 2: over-eager transpilers}

\subsubsection{Example 2a: vapourized $RZ$s}

\emph{Context}: The VQE circuits were initially implemented in PennyLane \cite{pennylane}. As the circuits were large it was desirable to heavily optimize them. Qiskit \cite{qiskit} 
has an extensive library of optimization passes, including 2-qubit gate resynthesis. PennyLane circuits were exported, optimized with the Qiskit transpiler, and imported back. \emph{Symptom:} The output of the VQE after automated circuit optimization was off by a few decimal points.
 
\emph{Analysis}: Hamiltonians consist of multiple Pauli terms, and their expectation values are measured term-by-term with circuits that perform basis rotations before measurement. In PennyLane a user only writes a single circuit, and others (with basis rotations) are generated and executed behind the scenes. Only the initial circuit was exported to the Qiskit transpiler, which removed $RZ$ gates at the end of the circuit, since they do not affect computational basis measurements. 
This optimized circuit, no longer equivalent to the original, was then used in  circuits with other measurements. 

This bug resulted from misalignment between the author's expectations of the transpiler and its behaviour. While EC would catch it, it is not always tractable, and it is not unreasonable to assume transpilation preserves equivalence, since generally it does.
Ultimately, the bug occurred because the author was working across frameworks that manage resources differently; developer-level knowledge of both frameworks (and compilation) was needed to diagnose and fix it.  
One workaround is to construct the full set of circuits and transpile them individually, but this would bottleneck larger workflows. Instead, the fix was easy but unsatisfying: inspect the circuits before and after transpilation for $RZ$ that got vapourized, and manually add them back. Developing  strategies for the average applications-level programmer who does not have in-depth knowledge of compilation pipelines is a larger problem.

\subsubsection{Example 2b: vapourized $CNOT$s} 
\emph{Context}: Error mitigation using zero-noise extrapolation (ZNE) \cite{zne} was performed to improve VQE hardware results. Circuits were generated to systematically scale up noise, and were executed on hardware with intent to extrapolate to the zero-noise limit. \emph{Symptom}: Results from the noisier circuits were not substantially different than the originals. 

\emph{Analysis}: In ZNE, systematic increase of noise is achieved by inserting \emph{redundant} gates that add noise but preserve circuit equivalence.
Due to higher two-qubit gate error rates, $CNOT$ pair insertion is often used: for every $CNOT$, noise is scaled by adding one extra pair of $CNOT$s, two extra pairs, etc. Suppose such circuits are generated and submitted to hardware. Unless they are already expressed using hardware-native gates, the hardware provider must translate them. 
A naive 1-1 translation will likely produce sub-optimal results, so it is of interest to the provider (and you, as the paying customer) to apply further optimizations. However, the inserted $CNOT$s do nothing, and a good compiler will vapourize them.
A user experiencing this may at first assume the system works well and error mitigation is not needed. However, the VQE circuits were large enough that increased noise should certainly be detected.

The issue here is that it's not a bug: \emph{it's a feature}, and not in the sarcastic sense of the phrase. 
Automated circuit optimization by the provider benefits the average user who is not familiar with the process.
Unlike the other bugs, a local debugging session would fail to identify the cause. EC or FV would not help, because the circuits are explicitly designed to be equivalent. The issue is challenging to identify without knowledge of the providers compilation stack. This could be avoided by allowing a user to fine-tune which parts of compilation are applied, at a high level, even if the details are hidden from the user.
In this case ``fix'' was bypassing the transpiler by submitting jobs in terms of native gates. However, the frameworks used did not have the requisite tools, so a separate transpiler needed to be written. This required working out decompositions, testing and EC, etc.
Ultimately, we needed a ``human compiler'', which introduces yet another source of errors; improving documentation for transpilers is essential for managing user expectations.

\section{Questions and concluding thoughts}

Following analysis of the studies, tools, and sample bugs, we propose two sets of research questions.
The first pertains to design of tools and frameworks that can help \emph{avoid} bugs like those in the examples. (1)~How can features such as EC and customization of compilation processes be integrated into frameworks to better meet user expectations, without them requiring in-depth knowledge? 
(2)~What visualization tools are effective for detecting and triggering quantum bugs? 
(3)~How do language features such as use of type systems or programming style (functional vs. object-oriented, etc.) affect the frequency, type, or location of bugs? 
(4)~How do we develop automated methods to analyze quantum programs and detect, raise warnings, and suggest fixes for quantum bugs?

The second set pertains more to how we reason about quantum programming. This article, and many tools and empirical studies, are biased towards a circuit-centric picture of algorithms written by developers with expert-level QC knowledge. But the gate model is not the only option, and building circuits gate-by-gate will not be practical for future large-scale algorithms. With that in mind, (1)~How do bug types and frequency vary between QC modalities (gate model, measurement-based, etc.), and are different debugging and testing strategies needed? (2)~How does bug type and frequency vary at different levels of abstraction, e.g., high-level languages where quantum aspects are largely hidden from users, vs. low-level systems working directly at the pulse level? (3)~What features can frameworks include to help teach quantum debugging to new users?

These questions will be challenging, but interesting to answer. Insight can be gained empirically by analyzing bug reports and case studies, and by adapting classical strategies. However, most require studying how people engage with quantum programming languages and tools.  
They are also complicated by the fact that the number of known quantum algorithms, and people implementing them, is very small. How will strategies change when algorithms don't have clear expressions to use for FV? When a program is too large to simulate, or to perform EC or FV, how will we even know something is wrong?
As QC gains traction, it will be critical to develop new tools and strategies for debugging. We hope this article spurs further interest in the area from the software community. Finally, we encourage you to \emph{share your mistakes} as done here, as they provide insight into our (mis)understanding of quantum algorithms, and serve to improve software for everyone.

\begin{acks}
ODM is funded by NSERC and the Canada Research Chairs program. Thanks to the participants of IEEE Quantum Week 2023 for inspiring discussions, the anonymous reviewers for their comments, and to Xanadu, where an early draft was written, for their hospitality.   
\end{acks}

\bibliographystyle{ACM-Reference-Format}
\bibliography{main}

\end{document}